\documentclass[conference]{IEEEtran}
\IEEEoverridecommandlockouts
\usepackage{cite}
\usepackage{amsmath,amssymb,amsfonts}
\usepackage{algorithmic}
\usepackage{graphicx}
\usepackage{textcomp}
\usepackage{xcolor}
\usepackage{booktabs}
\usepackage{multirow}
\usepackage{pifont}
\usepackage{array}
\usepackage{cite}
\usepackage{flushend}
\usepackage[percent]{overpic}
\usepackage[normalem]{ulem}
\def\BibTeX{{\rm B\kern-.05em{\sc i\kern-.025em b}\kern-.08em
    T\kern-.1667em\lower.7ex\hbox{E}\kern-.125emX}}
\newcommand{\xmark}{\ding{55}}

\DeclareMathAlphabet\mathbfcal{OMS}{cmsy}{b}{n}
\begin{document}

\title{On Speech Pre-emphasis as a Simple and Inexpensive Method to Boost Speech Enhancement\thanks{This work has been funded, in part, by the Spanish Ministry of Science and Innovation under the ``Ram\'on y Cajal'' programme (RYC2022-036755-I). Also in part, this work has been supported by the project PID2019-104206GB-I00 funded by MCIN/AEI/10.13039/501100011033.}
}

\author{\IEEEauthorblockN{Iv\'an L\'opez-Espejo$^1$, Aditya Joglekar$^2$, Antonio M. Peinado$^1$, Jesper Jensen$^{3,4}$}
\IEEEauthorblockA{\textit{$^1$Department of Signal Theory, Telematics and Communications, University of Granada, Spain}\\
\textit{$^2$Center for Robust Speech Systems (CRSS), The University of Texas at Dallas, USA}\\
\textit{$^3$Oticon A/S, Denmark}\\
\textit{$^4$Department of Electronic Systems, Aalborg University, Denmark}\\
\texttt{\{iloes,amp\}@ugr.es, aditya.joglekar@utdallas.edu, jesj@oticon.com}}
}

\maketitle

\begin{abstract}
Pre-emphasis filtering, compensating for the natural energy decay of speech at higher frequencies, has been considered as a common pre-processing step in a number of speech processing tasks over the years. In this work, we demonstrate, for the first time, that pre-emphasis filtering may also be used as a simple and computationally-inexpensive way to leverage deep neural network-based speech enhancement performance. Particularly, we look into pre-emphasizing the estimated and actual clean speech prior to loss calculation so that different speech frequency components better mirror their perceptual importance during the training phase. Experimental results on a noisy version of the TIMIT dataset show that integrating the pre-emphasis-based methodology at hand yields relative estimated speech quality improvements of up to 4.6\% and 3.4\% for noise types seen and unseen, respectively, during the training phase. Similar to the case of pre-emphasis being considered as a default pre-processing step in classical automatic speech recognition and speech coding systems, the pre-emphasis-based methodology analyzed in this article may potentially become a default add-on for modern speech enhancement.
\end{abstract}


\section{Introduction}
\label{sec:intro}

Speech is characterized by a spectral roll-off, or spectral tilt, stemming from glottal excitation due to vocal fold vibration \cite{Paavo17}. Consequently, more speech energy is found at lower frequencies than at higher ones. This spectral tilt may lead to speech processing systems ``overlooking'' higher frequencies \cite{Tom17}, a concern given that perceptually-relevant speech elements such as fricatives, affricates, and some plosives have higher energy at these frequencies \cite{Kent02,Monson12}.

The above issue has typically been addressed using pre-emphasis filtering, a simple yet effective speech pre-processing step that compensates high-frequency components by flattening the speech spectrum \cite{Tom17}. Although pre-emphasis filtering is a default consideration in classical automatic speech recognition (see, e.g., \cite{Marco11}) and speech coding systems \cite{Tom17}, its application and study have been minimal in the context of modern (i.e., deep neural network-based) speech enhancement. In \cite{Espejo23}, L\'opez-Espejo \emph{et al.} demonstrate that integrating pre-emphasis filtering into the scale-invariant signal-to-distortion ratio (SI-SDR) \cite{LeRoux19} loss function entails no advantage when this loss function is used to train an end-to-end speech enhancement system. Besides, the authors of the Speech Enhancement via Attention Masking Network (SEAMNET) \cite{SEAMNET} incorporate speech pre-emphasis into the mean squared error (MSE) loss function working in the time domain. However, they do not also look into an equivalent loss function without pre-emphasis, and, therefore, we are agnostic about any possible benefits that this type of filtering may bring for speech enhancement.

In contrast to \cite{Espejo23} and \cite{SEAMNET}, in this paper, we show that pre-emphasis filtering can be used as a simple and inexpensive method to boost modern speech enhancement. Specifically, we explore a straightforward integration, into the MSE loss function operating in the spectral magnitude domain, of two different pre-emphasis approaches: first-order high-pass finite impulse response (FIR) filtering \cite{Tom17} and equal-loudness pre-emphasis \cite{Hermansky90}. The modified loss function is employed to train a convolutional recurrent neural network (CRNN)-based speech enhancement system that follows a spectral masking approach \cite{JuanmaThesis}. Experimental results indicate that, compared to employing the standard MSE loss, incorporating pre-emphasis filtering and subsequent intensity-to-loudness conversion \cite{Hermansky90} results in relative improvements in speech quality\footnote{In this work, we consider the well-known perceptual evaluation of speech quality (PESQ) \cite{Rix01,PESQ} metric to test speech quality.} of up to 4.6\% and 3.4\% for noise types seen and unseen, respectively, during the training phase. It is particularly important to note that our work represents a first attempt successfully applying the pre-emphasis-based concept at hand, and its generalizability to other speech enhancement architectures/approaches and loss functions is subject of a future study.

The rest of this paper is structured as follows. Section \ref{sec:seframe} describes the speech enhancement framework considered in this work, while integration of speech pre-emphasis is explained in Section \ref{sec:preemphasis}. The speech dataset used for experimental purposes is detailed in Section \ref{sec:materials}. Results are discussed in Section \ref{sec:results}. Finally, Section \ref{sec:conclusion} concludes this work.


\section{Speech Enhancement Framework}
\label{sec:seframe}

\begin{figure*}[ht]
\centering
\begin{overpic}[width=\linewidth]{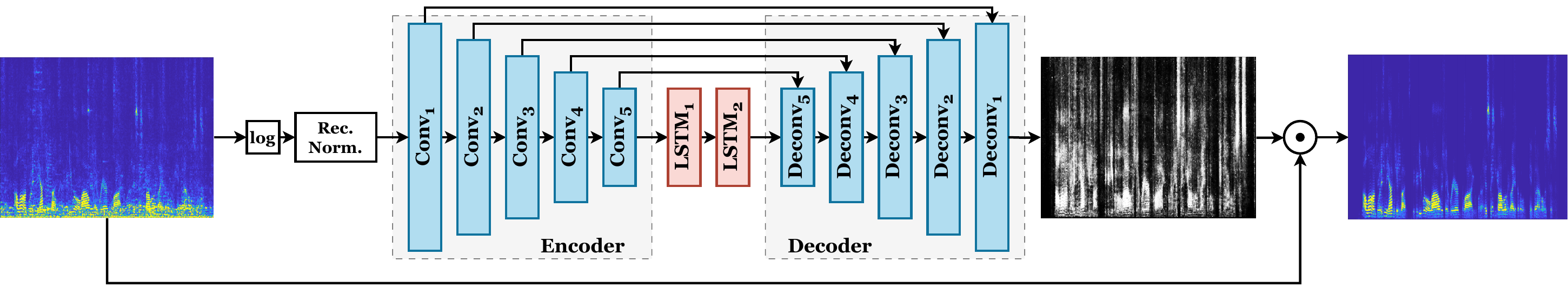}
\put(6.2,16.3){\scriptsize $\mathbf{Y}$}
\put(72,16.3){\scriptsize $\hat{\mathbfcal{M}}$}
\put(92.2,16.3){\scriptsize $\hat{\mathbf{X}}$}
\put(43.3,2.5){\scriptsize $\mathbf{f}\left(\cdot|\theta\right)$}
\end{overpic}
\caption{Block diagram of the speech enhancement system employed in this paper. ``Rec. Norm.'' stands for time-recursive mean normalization \cite{Jens18}. See the text for further details.}
\label{fig:sesystem}
\end{figure*}
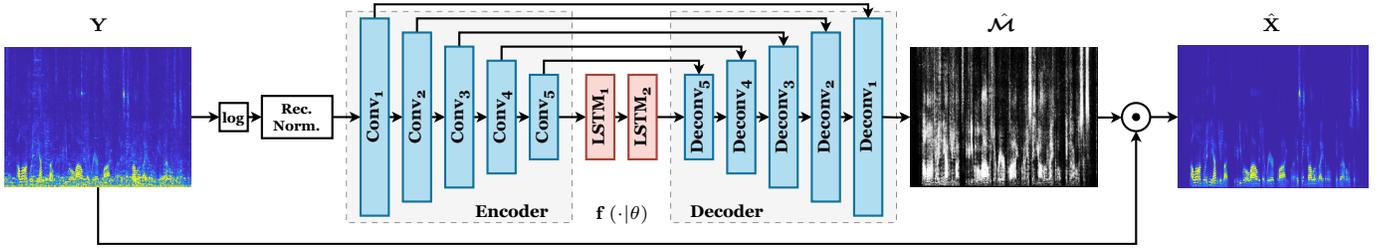

Let $y(m)$ be a (finite-length) time-domain noisy speech signal consisting of a distorted version of a clean speech signal $x(m)$ (for experimental purposes, we will consider in this work an additive noise distortion model, $y(m)=x(m)+\nu(m)$, where $\nu(m)$ is a background noise signal). This noisy signal can be expressed in the short-time Fourier transform (STFT) domain as $Y(k,t)$, where $k=0,...,K-1$ and $t=0,...,T-1$ denote the frequency bin and time frame indices, respectively. Besides, let
\begin{equation}
  \mathbf{Y}=\left[\begin{array}{ccc}
    |Y(0,0)| & \cdots & |Y(0,T-1)| \\
    \vdots & \ddots & \vdots \\
    |Y(K-1,0)| & \cdots & |Y(K-1,T-1)|
  \end{array}\right]
  \label{eq:y_spec}
\end{equation}
be a $K\times T$ matrix with the magnitude spectrum of $y(m)$.

Following the spectral masking approach illustrated by Fig. \ref{fig:sesystem}, our goal is to estimate the clean speech magnitude spectrum $\mathbf{X}$ (defined similarly to $\mathbf{Y}$) by means of a time-frequency mask $\hat{\mathbfcal{M}}\in[0, 1]^{K\times T}$. This mask is applied to $\mathbf{Y}$ via point-wise multiplication, specifically,
\begin{equation}
    \hat{\mathbf{X}}=\hat{\mathbfcal{M}}\odot\mathbf{Y},
    \label{eq:masking}
\end{equation}
where the $\odot$ operator denotes the Hadamard product, and $\hat{\cdot}$ means an estimate. To realize Eq. (\ref{eq:masking}), we aim at learning a mapping function $\mathbf{f}\left(\cdot|\theta\right):\mathbb{R}^{K\times T}\rightarrow[0, 1]^{K\times T}$ estimating $\hat{\mathbfcal{M}}$ from the noisy speech log-magnitude spectrum, after application of time-recursive mean normalization \cite{Jens18}, as in
\begin{equation}
  \hat{\mathbfcal{M}}=\mathbf{f}\left(\overline{\log\mathbf{Y}}|\theta\right),
  \label{eq:mask}
\end{equation}
where $\theta$ is the set of learnable parameters of the mapping function, the log operator is applied element-wise, and $\overline{\cdot}$ denotes time-recursive mean normalization.

The mapping function $\mathbf{f}(\cdot|\theta)$ is deployed by the CRNN depicted in Fig. \ref{fig:sesystem} \cite{JuanmaThesis}. This architecture is comprised of an encoder with 5 convolutional layers followed by 2 long short-term memory (LSTM) layers and a decoder with 5 deconvolutional layers. All the convolutional and deconvolutional layers employ $3\times 1$ kernels, a stride of $(2, 1)$, and exponential linear unit (ELU) activations (except for the output layer, which uses a sigmoid activation function). The $i$-th convolutional layer, \texttt{Conv$_i$}, has $2^{i+2}$ feature maps. Similarly, the $i$-th deconvolutional layer, \texttt{Deconv$_i$}, has $2^{i+1}$ feature maps (except for \texttt{Deconv$_1$}, which has 1 only). A skip connection serves to concatenate the output of \texttt{Conv$_i$} to the input of \texttt{Deconv$_i$}. The LSTM hidden state dimension is set to 1,024.

Once $\hat{\mathbf{X}}$ has been obtained from Eq. (\ref{eq:masking}), the enhanced speech waveform $\hat{x}(m)$ is synthesized by calculating the inverse STFT of $\hat{X}(k,t)=\left|\hat{X}(k,t)\right|\cdot e^{j\measuredangle Y(k,t)}$, where the symbol $\measuredangle$ denotes the phase value of the STFT coefficient.

It should be noted that recent speech enhancement efforts (e.g., \cite{Bu22,Close23}) employ spectral masking schemes similar to the one described in this section.

\subsection{Implementation Details}
\label{ssec:issues}

For STFT computation, we make use of a Hann window with a length of 32 ms and a shift of 16 ms. Moreover, the total number of frequency bins is $K=257$.

Using Adam \cite{Adam} with default parameters, the deep neural network parameter set $\theta$ is optimized towards minimizing the MSE between estimated and actual training clean speech magnitude spectra:
\begin{equation}
    \mathcal{L}_{\mbox{\scriptsize MSE}}=\frac{1}{KT}\sum_{k=0}^{K-1}\sum_{t=0}^{T-1}\left(\left|\hat{X}(k,t)\right|-\left|X(k,t)\right|\right)^2.
    \label{eq:mse_loss}
\end{equation}
In addition, the mini-batch size is 8 training utterances, early-stopping \cite{Gershenfeld88} with a patience of 15 epochs is employed, and training runs for a maximum of 200 epochs.

\section{Speech Pre-emphasis Integration}
\label{sec:preemphasis}

We explore methods for pre-emphasizing the estimated and actual training clean speech during deep neural network training so that speech is perceptually balanced prior to loss calculation \cite{SEAMNET,Espejo23}. By doing this, our expectation is that the contribution of distinct speech frequency components to the total loss better reflects their perceptual importance, thus boosting speech enhancement performance.

We consider two pre-emphasis variants that can be easily integrated into the speech enhancement loss function: standard pre-emphasis consisting of a first-order high-pass FIR filtering (Subsec. \ref{ssec:standard}), and equal-loudness pre-emphasis (Subsec. \ref{ssec:equal-loudness}). Besides, cubic-root amplitude compression is optionally considered to leverage pre-emphasis by further reducing the speech spectral magnitude variation (Subsec. \ref{ssec:3root}).

Although the formulae below are particularized to the MSE loss function of Eq. (\ref{eq:mse_loss}), it is important to note that the pre-emphasis-based methodology under consideration can, in principle, be adapted to any speech enhancement loss function. Hence, this simple and cheap methodology may potentially become a \emph{default add-on} for training deep neural network-based speech enhancement systems.

\subsection{Standard Speech Pre-emphasis (SP)}
\label{ssec:standard}

Standard speech pre-emphasis is implemented by a first-order high-pass FIR filter \cite{Tom17}, whose magnitude response is
\begin{equation}
  \begin{array}{lll}
    \left|H_{\mbox{\scriptsize SP}}(f)\right| & = & \left|1-\alpha e^{-j2\pi f/f_s}\right| \vspace{0.25cm} \\
    & = & \sqrt{\alpha^2-2\alpha\cos(2\pi f/f_s)+1},
  \end{array}
  \label{eq:sp}
\end{equation}
where $f$ and $f_s$ denote frequency and the sampling rate, respectively, in Hz, and $0<\alpha<1$ is a free parameter controlling the sharpness of the filter response (see Fig. \ref{fig:filters}).

Let $\left|\bar{H}_{\mbox{\scriptsize SP}}(f)\right|\in(0, 1]$ represent a scaled version of Eq. (\ref{eq:sp}) that is obtained by normalizing $\left|H_{\mbox{\scriptsize SP}}(f)\right|$ to have a maximum amplitude of 1. Then, $\left|\bar{H}_{\mbox{\scriptsize SP}}(k)\right|$ is found by uniform sampling of $\left|\bar{H}_{\mbox{\scriptsize SP}}(f)\right|$, and the former quantity is used to compute pre-emphasized versions of the estimated and actual clean speech magnitude spectra, respectively, as follows:
\begin{equation}
    \begin{array}{ccc}
    \left|\hat{X}_{\mbox{\scriptsize SP}}(k,t)\right| & = & \left|\bar{H}_{\mbox{\scriptsize SP}}(k)\right|\cdot\left|\hat{X}(k,t)\right|, \vspace{0.25cm} \\
    \left|X_{\mbox{\scriptsize SP}}(k,t)\right| & = & \left|\bar{H}_{\mbox{\scriptsize SP}}(k)\right|\cdot\left|X(k,t)\right|.
    \end{array}
    \label{eq:filtering}
\end{equation}
Finally, $\left|\hat{X}_{\mbox{\scriptsize SP}}(k,t)\right|$ and $\left|X_{\mbox{\scriptsize SP}}(k,t)\right|$ are employed to replace $\left|\hat{X}(k,t)\right|$ and $\left|X(k,t)\right|$, respectively, in the MSE loss function of Eq. (\ref{eq:mse_loss}), $\mathcal{L}_{\mbox{\scriptsize MSE}}$.

\subsection{Equal-loudness Pre-emphasis (ELP)}
\label{ssec:equal-loudness}

\begin{figure}[t]
\centering
\includegraphics[width=0.9\linewidth]{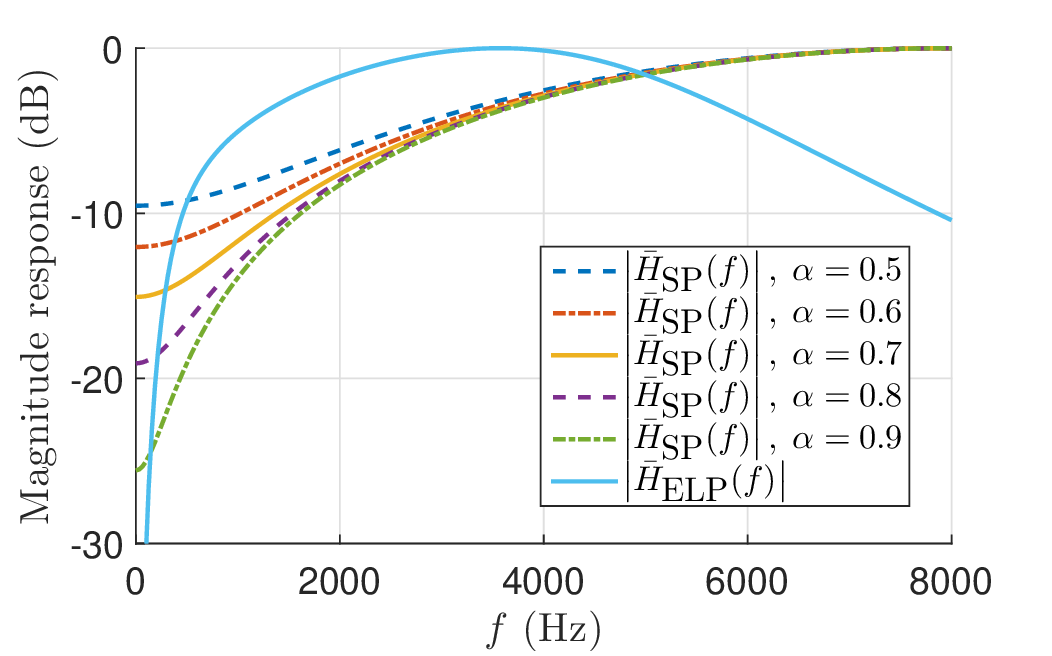}
\caption{A comparison between the normalized magnitude responses of standard pre-emphasis (given various values of $\alpha$) and equal-loudness pre-emphasis.}
\label{fig:filters}
\end{figure}

As an alternative to standard speech pre-emphasis, equal-loudness pre-emphasis, proposed by Hermansky \cite{Hermansky90} and accounting for the psychophysics of hearing, may be used. The equal-loudness pre-emphasis magnitude response, $|H_{\mbox{\scriptsize ELP}}(f)|$, approximates the frequency-dependent sensitivity of human hearing at about the 40 dB level:
\begin{equation}
  |H_{\mbox{\scriptsize ELP}}(f)|=\sqrt{\frac{\left(f^2+\beta_1\right)f^4}{\left(f^2+\beta_2\right)^2\left(f^2+\beta_3\right)\left((2\pi f)^6+\beta_4\right)}},
\end{equation}
where $\beta_1=1.44\cdot10^6$, $\beta_2=1.6\cdot10^5$, $\beta_3=9.61\cdot10^6$, and $\beta_4=9.58\cdot10^{26}$.

A procedure similar to that of the previous subsection is then followed. First, $|H_{\mbox{\scriptsize ELP}}(f)|$ is scaled to have a maximum amplitude of 1 and produce $|\bar{H}_{\mbox{\scriptsize ELP}}(f)|\in[0, 1]$, which, in turn, is uniformly sampled to obtain $|\bar{H}_{\mbox{\scriptsize ELP}}(k)|$. Second, the latter quantity is applied as in Eq. (\ref{eq:filtering}) to calculate $\left|\hat{X}_{\mbox{\scriptsize ELP}}(k,t)\right|$ and $\left|X_{\mbox{\scriptsize ELP}}(k,t)\right|$, which are used to replace $\left|\hat{X}(k,t)\right|$ and $\left|X(k,t)\right|$, respectively, in Eq. (\ref{eq:mse_loss}).

Fig. \ref{fig:filters} shows a comparison between the normalized magnitude responses of standard pre-emphasis ---given several values of $\alpha$--- and equal-loudness pre-emphasis. As can be seen, unlike standard speech pre-emphasis, equal-loudness pre-emphasis accounts for the decrease in hearing sensitivity at higher frequencies \cite{Hermansky90,Florian05}.

\subsection{Intensity-to-loudness Conversion (I2L)}
\label{ssec:3root}

In his pipeline definition of the well-known perceptual linear prediction (PLP) acoustic features \cite{Hermansky90}, Hermansky incorporated cubic-root amplitude compression after equal-loudness pre-emphasis to simulate the non-linear relationship between the intensity of sound and its perceived loudness \cite{Stevens57}. Motivated by this, we optionally consider cubic-root amplitude compression by applying the operator $(\cdot)^{2/3}$ to the pre-emphasized versions (\emph{regardless of the pre-emphasis type}) of the estimated and actual clean speech magnitude spectra before they are used in the MSE loss function of Eq. (\ref{eq:mse_loss}). Notice that cubic-root amplitude compression can boost the effect of pre-emphasis by further reducing the dynamic range of the speech magnitude spectrum.

\section{Speech Dataset}
\label{sec:materials}

For experimental purposes, we use the TIMIT-1C speech dataset \cite{JuanmaThesis}, which is comprised of clean and simulated noisy speech signals at a sampling rate of 16 kHz. First, 300 clean speech signals were created from the speaker-wise concatenation of utterances from the well-known TIMIT dataset \cite{TIMIT,Lamel89} for the clean speech signals to have a duration between 6 and 10 seconds. Second, these clean speech signals were artificially distorted by diverse types of additive noise at distinct signal-to-noise ratios (SNRs) to generate simulated noisy speech samples.

TIMIT-1C is composed of three sets: training, validation and test, to which 200/300, 50/300 and 50/300 unique clean speech signals, respectively, are assigned. The training and validation sets contain speech signals degraded by noise types ``car'', ``bus station'', ``restaurant'', and ``street''. The test set includes speech signals distorted by, in addition to the previous noises (\emph{seen noises}), the noise types ``caf\'e'', ``train station'', ``pedestrian street'', and ``bus'' (\emph{unseen noises}). The three sets consider the same discrete set of SNRs: $\{-5, 0, 5, 10, 15, 20\}$ dB. Neither noise realizations nor speakers overlap across sets, and the number of female and male speakers in each set is balanced. In total, the training, validation and test sets are composed of
\begin{enumerate}
    \item $200\;\mbox{clean signals}\times 4\;\mbox{noises}\times 6\;\mbox{SNRs}=4,800$,
    \item $50\;\mbox{clean signals}\times 4\;\mbox{noises}\times 6\;\mbox{SNRs}=1,200$, and
    \item $50\;\mbox{clean signals}\times 8\;\mbox{noises}\times 6\;\mbox{SNRs}=2,400$
\end{enumerate}
noisy speech samples, respectively \cite{JuanmaThesis}.

\begin{table*}[t]
	\centering
	\setlength{\extrarowheight}{3pt}
    \caption{Speech enhancement results in terms of STOI and PESQ. Results are broken down by SNR and seen/unseen noises. SP, ELP and I2L stand for standard pre-emphasis, equal-loudness pre-emphasis and intensity-to-loudness conversion, respectively. The symbol \xmark\; means that pre-emphasis or intensity-to-loudness conversion is not applied. Best PESQ results are marked in boldface. See the text for further details.}
	\resizebox*{!}{\columnwidth}{\begin{tabular}{cc|c|c|cc|cc|c|c|cc|cc}
			\toprule[1pt]\midrule[0.3pt]
			\multicolumn{1}{c}{\textbf{SNR}} & \textbf{Metric} & \multicolumn{6}{c|}{\textbf{Seen noises}} & \multicolumn{6}{c}{\textbf{Unseen noises}} \\ \cline{3-14}
            \textbf{(dB)} & & \emph{Noisy} & \multicolumn{5}{c|}{\emph{Processed}} & \emph{Noisy} & \multicolumn{5}{c}{\emph{Processed}} \\ \cline{3-14}
            & & & \xmark & \multicolumn{2}{c|}{+SP} & \multicolumn{2}{c|}{+ELP} & & \xmark & \multicolumn{2}{c|}{+SP} & \multicolumn{2}{c}{+ELP} \\ \cline{4-8} \cline{10-14}
			\multicolumn{1}{c}{} & & & \xmark & \xmark & +I2L & \xmark & +I2L & & \xmark & \xmark & +I2L & \xmark & +I2L \\
			\midrule[0.1pt]\midrule[0.1pt]
			\multirow{2}{*}{-5} & STOI & 0.64 & 0.74 & 0.74 & 0.74 & 0.74 & 0.74 & 0.65 & 0.73 & 0.73 & 0.73 & 0.73 & 0.73 \\
			& PESQ & 1.06 & 1.57 & 1.59 & \textbf{1.62} & 1.50 & 1.58 & 1.16 & 1.47 & 1.47 & \textbf{1.49} & 1.47 & 1.48 \\
			\midrule
			\multirow{2}{*}{0} & STOI & 0.73 & 0.84 & 0.84 & 0.84 & 0.83 & 0.83 & 0.75 & 0.83 & 0.83 & 0.83 & 0.83 & 0.83 \\
			& PESQ & 1.11 & 1.86 & 1.90 & \textbf{1.93} & 1.81 & 1.89 & 1.27 & 1.76 & 1.76 & \textbf{1.81} & 1.78 & 1.78 \\
			\midrule
			\multirow{2}{*}{5} & STOI & 0.82 & 0.90 & 0.90 & 0.90 & 0.90 & 0.90 & 0.83 & 0.90 & 0.90 & 0.90 & 0.90 & 0.90 \\
			& PESQ & 1.25 & 2.20 & 2.26 & \textbf{2.31} & 2.21 & 2.23 & 1.51 & 2.14 & 2.15 & \textbf{2.21} & 2.20 & 2.17 \\
			\midrule
			\multirow{2}{*}{10} & STOI & 0.89 & 0.94 & 0.94 & 0.94 & 0.94 & 0.94 & 0.91 & 0.94 & 0.95 & 0.95 & 0.94 & 0.94 \\
			& PESQ & 1.53 & 2.61 & 2.67 & \textbf{2.72} & 2.65 & 2.64 & 1.84 & 2.56 & 2.59 & \textbf{2.66} & \textbf{2.66} & 2.62 \\
			\midrule
			\multirow{2}{*}{15} & STOI & 0.94 & 0.97 & 0.97 & 0.97 & 0.97 & 0.97 & 0.95 & 0.97 & 0.97 & 0.97 & 0.97 & 0.97 \\
			& PESQ & 1.92 & 2.94 & 3.00 & \textbf{3.08} & 3.02 & 3.01 & 2.26 & 2.93 & 2.96 & \textbf{3.04} & \textbf{3.04} & 3.00 \\
			\midrule
			\multirow{2}{*}{20} & STOI & 0.97 & 0.98 & 0.98 & 0.98 & 0.98 & 0.98 & 0.98 & 0.98 & 0.98 & 0.98 & 0.98 & 0.98 \\
			& PESQ & 2.45 & 3.30 & 3.35 & \textbf{3.45} & 3.38 & 3.38 & 2.84 & 3.32 & 3.37 & \textbf{3.44} & 3.43 & 3.40 \\
			\midrule[0.3pt]\bottomrule[1pt]
	\end{tabular}}
	\label{tab:results}
\end{table*}

\section{Experimental Results}
\label{sec:results}

In this section, we evaluate the pre-emphasis-based procedures presented in Section \ref{sec:preemphasis} in terms of estimated quality and intelligibility of the enhanced speech by means of PESQ (Perceptual Evaluation of Speech Quality) \cite{Rix01,PESQ} and STOI (Short-Time Objective Intelligibility) \cite{Taal11}, respectively.

First of all, it is important to point out that preliminary experiments revealed that, when considering standard speech pre-emphasis (see Subsec. \ref{ssec:standard}), $\alpha=0.6$ is a good choice, and, therefore, we use this parameter value in the rest of this section. That being said, we also observed that the value of $\alpha$ has a relatively low impact on speech enhancement performance as long as it is not too close to either 0 or 1.

Table \ref{tab:results} displays STOI and PESQ results calculated from speech signals processed by the speech enhancement system of Section \ref{sec:seframe} when this system integrates no pre-emphasis (\emph{baseline system}), standard pre-emphasis (+SP) or equal-loudness pre-emphasis (+ELP). In case pre-emphasis is integrated, results are broken down by whether (+I2L) or not intensity-to-loudness conversion is applied. As a reference, STOI and PESQ scores computed from the original noisy (namely, unprocessed) speech signals are also shown. All the results are broken down by SNR and seen/unseen noises. Note that, in Table \ref{tab:results}, the symbol \xmark\; means that pre-emphasis or intensity-to-loudness conversion is not applied.

On the one hand, we can see from Table \ref{tab:results} that, according to STOI scores, pre-emphasis filtering has no impact on speech intelligibility. On the other hand, we can also see from this table that the integration of pre-emphasis yields, with respect to the baseline system, equal or better speech quality (PESQ) results for all the noisy conditions evaluated except for only seen noises at -5 dB and 0 dB SNRs when equal-loudness pre-emphasis is considered. In particular, the best speech quality results are obtained when standard pre-emphasis is integrated into the MSE loss function $\mathcal{L}_{\mbox{\scriptsize MSE}}$ and intensity-to-loudness conversion follows. On average, this approach achieves PESQ relative improvements over the baseline of around 4.6\% and 3.4\% for seen and unseen noises, respectively.

The above results indicate that perceptually balancing the estimated and actual clean speech signals prior to loss calculation allows for obtaining supplementary speech quality gains over a conventionally-trained modern speech enhancement system. Moreover, a major advantage of adopting the proposed strategy to improve speech quality is the minimal additional computational cost at training time, and no additional cost at inference time. Altogether, these findings suggest that the pre-emphasis-based methodology studied here may be potentially adopted as a default add-on in modern speech enhancement, similar to the case of pre-emphasis filtering being embraced as a default pre-processing step in other kinds of speech processing systems like classical automatic speech recognition and speech coding systems.

\section{Conclusion}
\label{sec:conclusion}

To the best of our knowledge, the present work constitutes the first successful attempt to empirically demonstrate that deep neural network-based speech enhancement performance can be boosted ---in a straightforward and computationally-inexpensive manner--- through the integration of pre-emphasis filtering. A future survey will investigate the generalizability of the pre-emphasis-based methodology studied in this paper by exploring multiple speech enhancement architectures/approaches and loss functions that may operate in different signal domains. Furthermore, such a future survey will also look into running subjective listening tests to contrast what is predicted by objective speech quality and intelligibility metrics in order to strengthen the conclusions drawn.

\section*{Acknowledgement}

We would like to thank our student Mr. Antonio Garrido Molina for partly contributing to the experimental development of the present work.

\bibliographystyle{IEEEbib}
\bibliography{strings,refs}

\end{document}